\documentclass[showpacs,preprint]{revtex4}
\usepackage{graphicx}
\usepackage{dcolumn}
\usepackage{amsmath}
\usepackage[latin1]{inputenc}
\usepackage{graphicx, psfrag}
\usepackage{amssymb}
\usepackage[colorlinks=true, citecolor=blue, urlcolor = blue, linkcolor= red, bookmarks=true]{hyperref}
\usepackage{float}
\usepackage{amsmath}
\usepackage{amsfonts}
\usepackage{dcolumn}
\usepackage{hyperref}
\usepackage{subfigure}
\usepackage{pgfplots}
\usepackage{epstopdf}
\usepackage{booktabs}

\makeatletter
\def\btt#1{\texttt{\@backslashchar#1}}
\DeclareRobustCommand\bblash{\btt{\@backslashchar}} \makeatother

\makeatletter
\def\btt#1{\texttt{\@backslashchar#1}}
\DeclareRobustCommand\bblash{\btt{\@backslashchar}} \makeatother

\begin{document}
\title{Rotating regular black holes in AdS spacetime and its shadow}
 
\author{Balendra Pratap Singh$^{a}$}\email{balendra29@gmail.com}
\author{Md Sabir Ali$^{b}$}\email{alimd.sabir3@gmail.com}

\author{Sushant~G.~Ghosh$^{c;d}$} \email{sghosh2@jmi.ac.in, sgghosh@gmail.com}
\affiliation{$^{a}$ Tulas Institute, Dehradun, Uttarakhand 248197, India}
\affiliation{$^{b}$ Indian Institute of Science Education and Research Kolkata, West Bengal 741246, India }
\affiliation{$^{c}$ Centre for Theoretical Physics, Jamia Millia Islamia, New Delhi 110025, India}

\affiliation{$^{d}$ Astrophysics and Cosmology
Research Unit, School of Mathematics, Statistics and Computer Science, University of
KwaZulu-Natal, Private Bag 54001, Durban 4000, South Africa}

\begin{abstract}

The presence of a photon region around the black hole is an essential feature in receiving the emitted spectrum from the vicinity of the black hole by the distant observer. In this paper, we investigate the optical properties of rotating regular anti-de Sitter (AdS) black holes, which characterized by its mass $(M)$, spin parameter $(a)$, deviation parameter $(k)$ and the cosmological constant $(\Lambda)$ related to the curvature radius via $(\Lambda=-3/l^2)$.  We derive the complete null geodesic equations of motion and study the unstable circular orbits for an observer at given Boyer-Lindquist coordinates ($r_O,\;\vartheta_O$) in the domain of outer communication and trace the photon rings. For the analytical part of our study, we investigate the observables, namely, shadow radius $(R_s)$ and distortion parameter $(\delta_s)$. 
These rotating regular AdS black holes have smaller shadows when compare to the Kerr and regular spacetimes. We also estimate the energy emission rate of the black hole.
\end{abstract}
\maketitle
\section{Introduction}
Observing the rotating black hole shadow is one of the possible method to measure the black hole parameters. The incoming photons from the source having larger angular momentum get scattered from the black hole and reach the faraway observer and those having smaller angular momentum fall inside the black hole and form a dark shadow region. The black hole shadow forms in the close vicinity of the event horizon and thereby helps us to have the idea of the basic geometrical structure of horizons.  Recent astronomical observations from Event Horizon Telescope (EHT) confirm the existence of the black hole Sagittarius A* at the center of our galaxy Milky way \cite{Akiyama:2022L12,Akiyama:2022L13,Akiyama:2022L14,Akiyama:2022L15,Akiyama:2022L16,Akiyama:2022L17} and a supermassive black hole at the galactic center of Messier 87 \cite{Akiyama:2019cqa,Akiyama:2019brx,Akiyama:2019sww,Akiyama:2019bqs,Akiyama:2019fyp,Akiyama:2019eap}.  The observed image of the black hole is a two-dimensional dark disk surrounded by luminous rings. The dark region is  black hole shadow and the luminous rings are photon orbits.  

The apparent shape of the shadow also depends on the matter fields surrounding the black hole. Hence, the shadow of the black hole is treated as a useful tool to detect black hole spin and other deformation parameters because its shape and size carry the fingerprint of the geometry around the black hole. 
The astrophysical black hole candidates are having an appreciable amount of rotation. This consideration motivates us to study the rotating black holes and the shadow cast by them. Since the regular black holes contain Kerr-metric as a special case, we consider them as a prototype for a large class of non-Kerr families where the metric would have the same form with mass parameter $M$ simply replaced by $m(r)$. The analysis of the shadows will be a useful tool for better understanding of astrophysical black holes and also for comparing modified theories of gravity (MGs) with general theory of gravity (GR). In GR the black hole shadow is completely characterized by its Mass $M$ and spin parameter $a$ \cite{Bardeen:1973gb,Young:1976} but in MGs, there are also some additional free parameters which affect the shape and size of the black hole shadow. Hence, it is useful to study further the shadow of black holes in various available MGs. In the last few years, the study of black hole shadow in various spacetimes has been investigated \cite{Johannsen:2015qca,De,Amarilla:2010zq,Amarilla:2013sj,Yumoto:2012kz,Abdujabbarov:2016hnw,Amir:2016cen,Tsukamoto:2014tja,Hioki:2009na,Bambi:2010hf,Goddi:2016jrs,Takahashi:2005hy,Wei:2013kza,Abdujabbarov:2012bn,Amarilla:2011fx,Bambi:2008jg,Atamurotov:2013sca,Wang:2017hjl,Schee:2008kz,Grenzebach:2014fha,Singh:2017xle,Kumar:2019ohr,Kumar:2017tdw,Kumar:2020owy,Narzilloev:2022bbs,Wei:2019pjf,Papnoi:2021rvw,Atamurotov:2021hck}.  The effect of the plasma environment on the black hole shadow has been also discussed in \cite{Perlick:2015vta,Atamurotov:2015nra,Abdujabbarov:2015pqp}. The exotic feature of the black hole's shadow has also been explored in the higher-dimensional spacetime \cite{Papnoi:2014aaa,Singh:2017vfr,Amir:2017slq,Abdujabbarov:2015rqa,Atamurotov:2021cgh}.   Cunha et. al. \cite{Cunha:2015yba,Cunha:2016bpi} developed a ray tracing method for a better understanding of photon rays in the vicinity of Kerr spacetime with and without  scalar hair. 

Here, we shall adopt the formalism presented in \cite{Grenzebach:2014fha} for an observer located at some given Boyer-Lindquist coordinates ($r_O,\;\vartheta_O$) instead of infinity and study the shadow images of the regular anti-de Sitter (AdS) black holes. The shadow images for Kerr-Newman-NUT-AdS black holes and rotating AdS black holes in the braneworld scenario have been studied \cite{Grenzebach:2014fha,Eiroa:2017uuq}. Belhaj et. al. \cite{Belhaj:2021rae,Belhaj:2022rmc}, investigated   shadows of superentropic and Bardeen black holes in AdS spacetime. Shadows of rotating black holes with cosmological constant have been studied in \cite{Belhaj:2020kwv}. The authors of \cite{Afrin:2021ggx}, estimated the cosmological constant from Kerr-de sitter black holes.

In this work, we analytically estimate the effects of the spin and the related deformation parameters on the size and shape of the rotating regular AdS black hole. The size of the black hole shadow decreases and gets more distorted in comparison with the Kerr \cite{Bardeen} and nonsingular \cite{Amir:2016cen} black hole in AdS spacetime.

The paper is arranged as follows: In Sec.~\ref{sect2}, we describe the rotating regular AdS black hole. The geodesics structure of a test particle in this spacetime has been discussed in Sec.~\ref{sect3}. Further, in Sec.~\ref{sect4}, we derive an analytical formula for the shadow and plot them for different values of black hole parameters. In Sec.~\ref{sect5}, we estimate the energy emission rate and  finally, we conclude our results in Sec.~\ref{sect6}.
\section{Rotating regular ads black hole}\label{sect2} 
In this section, we study the general solution to   static spherically symmetric regular black holes and their rotating counterparts in AdS spacetime due to the presence of nonlinear electrodynamics. We start with a four-dimensional static black hole with spherical symmetry described by
\begin{eqnarray}\label{metric}
ds^2&=&-f(r)dt^2+\frac{dr^2}{f(r)}+r^2d\Omega_{2}^2,
\end{eqnarray}
where $f(r)$ is the metric function and $d\Omega_{2}=d\theta^2+\sin^2\theta d\phi^2$ is the line element on a two-dimensional unit sphere. The action for regular AdS spacetimes has the following form 
\begin{eqnarray}
\mathcal{I}=\int{d^4x}\sqrt{-g}[R-\mathcal{L\left(F\right)}+6\l^{-2}],
\end{eqnarray} 
where $R$ is the Ricci scalar, $\l$ is the curvature radius related to the cosmological constant $\Lambda=-3/\l^2$, and $\mathcal{L\left(F\right)}$ is the Lagrangian of the non-linear electrodynamics source having  the form 
\begin{eqnarray}\label{lagran}
\mathcal{L\left(F\right)}= e^{\left[{-s\left(2g^2\mathcal{F}\right)^{\frac{1}{4}}}\right]}\mathcal{F},
\end{eqnarray}
with $\mathcal{F}=F_{\mu\nu}F^{\mu\nu}/4$, where $F_{\mu\nu}=2\nabla_{[\mu}A_{\nu]}$. We take $s=q/2M$, where $q$ and $M$ correspond to the charge and the mass of the regular black hole. The variation of the action $\mathcal{I}$, with respect to $g_{\mu\nu}$ and $A_{\mu}$ leads to the field equations 
\begin{eqnarray}\label{fieldeqs1}
G_{\mu\nu}-\frac{3}{\l^2}=T_{\mu\nu},\;\;\; \nabla_{\mu}\left(\frac{\partial \mathcal{L\left(F\right)}}{\partial{F}}{F^{\nu\mu}}\right)=0,
\end{eqnarray}
where $T_{\mu\nu}$ is energy-momentum tensor given as
\begin{eqnarray}
\label{Tmunu}
T_{\mu\nu}=2\left[\frac{\partial\mathcal{L\left(F\right)}}{\partial{F}}F_{\mu\alpha}F_{\nu}^{\alpha}-g_{\mu\nu}\mathcal{L\left(F\right)}\right],
\end{eqnarray}
The form of $F_{\mu\nu}$ for a nonlinear charged follows the $\textit{ansatz}$
\begin{eqnarray}\label{Fmunu}
F_{\mu\nu}=2\delta^{\theta}_{[\mu}\delta^{\phi}_{\nu]}q\left(r\right)\sin\theta.
\end{eqnarray}
Eq.~(\ref{fieldeqs1})  implies $dF=0$ , which in turn reads $q^\prime(r)dr\wedge d\theta\wedge  d\phi=0$, leading to $q(r)=constant=q$. The constant $q$ is identified as nonlinear charge arising from nonlinear electrodynamics. Therefore, the resulting Maxwell tensor has the only one component
\begin{eqnarray}\label{maxwell}
F_{\theta\phi}=2q\sin\theta,
\end{eqnarray}
and 
\begin{eqnarray}
\mathcal{F}=\frac{q^2}{2r^4},
\end{eqnarray}
The Lagrangian (\ref{lagran}) of the nonlinear electrodynamics thus reads
\begin{eqnarray}\label{lagran1}
\mathcal{L\left(F\right)}=\frac{q^2}{2r^4}e^{-\frac{q^2}{2M r}}.
\end{eqnarray}
The $(t,t)$-component of the field equations (\ref{fieldeqs1}) along with the energy-momentum tensor (\ref{Tmunu}) and the Lagrangian (\ref{lagran1}) leads the metric function with $q^2=2Mk$, having the form
\begin{eqnarray}
\label{metricf}
f(r)=1-\frac{2M\;e^{{-\frac{k}{r}}}}{r}+\frac{r^2}{\l^2}.
\end{eqnarray}
In summary the metric for a non-singular AdS black holes in $4$-dimensional spacetime reads
\begin{eqnarray}\label{metric2}
\mathrm{ds^2}=-\left(1-\frac{2M\;e^{{-\frac{k}{r}}}}{r}+\frac{r^2}{\l^2}\right)dt^2+\frac{1}{\left(1-\frac{2M\;e^{{-\frac{k}{r}}}}{r}+\frac{r^2}{\l^2}\right)}dr^2+r^2 d\Omega_{2}^2
\end{eqnarray}
where $M\;\text{and}\;k$ are, respectively, the Arnowitt-Deser-Misner (ADM) mass and the nonlinear parameter of the black hole.
Next, we turn to discuss the rotating counterpart of the regular AdS black hole (\ref{metric2}). Before going into the main discussion, we write the gauge potential for a magnetically charged black hole. The electromagnetic potential for the magnetically charged rotating black hole was recently studied and derived using the usual Newman-Janis procedure, which leads to having the form \cite{Toshmatov:2017zpr, Erbin:2016lzq}
\begin{eqnarray}
\label{rot_gauge}
A_\mu=-\frac{g\;a\;\cos\theta}{\Sigma}\delta_\mu^t+\frac{g(r^2+a^2)\cos\theta}{\Sigma}\delta_\mu^\phi.
\end{eqnarray}
The gauge potential in the absence of the rotation ($a=0$), reduces to the expression for spherically symmetric spacetime is written as $A_\mu=g\cos\theta\delta_\mu^\phi$.
With this gauge potential in hand, we have Maxwell field invariant to be \cite{Toshmatov:2017zpr}
\begin{eqnarray}
\label{lagran_rot}
\mathcal{F}=\frac{2g^2\left(\left(\Sigma-2r^2\right)^2-4a^2 r^2\cos^2\theta\right)}{\Sigma^4}.
\end{eqnarray}
The Einstein's field equations lead to have the expressions of Lagrangian  \cite{Toshmatov:2017zpr} in the form
\begin{eqnarray}
\label{lag_rot}
\mathcal{L}(r)&=&\frac{4r^2\left(2r^2\Sigma\;m^{\prime\prime}+5\;m^\prime\left(\left(\Sigma-2r^2\right)^2-8r^2\Sigma+4r^2\right)\right)}{\Sigma^4},\nonumber\\
\mathcal{L}_\mathcal{F}(r)&=&\frac{r\Sigma\;m^{\prime\prime}+2m^\prime\left(\Sigma-2r^2\right)}{2g^2},
\end{eqnarray}
where $^{\prime}$ and $^{\prime\prime}$ denotes, respectively, the first and second derivatives with respect to the radial coordinate $r$.
The rotating regular black holes are described by four parameters, e.g., the mass $M$, the rotation parameter $a$, the nonlinear parameter $q$. The rotating regular metric belongs to prototype family of non-Kerr solutions, which in Boyer-Lindquist coordinates has same form as that of Kerr metric with mass $m$ replaced by some mass function $\tilde{m}$ which contains an additional deviation parameter $q$  from the NED.
 The regular rotating AdS black holes are stationary and axially symmetric spacetimes with a negative cosmological constant. The exact solution of the Einstein field equations with a negative cosmological constant that describes the rotating black holes in four-dimensional spacetime with asymptotic AdS behavior in the standard Boyer-Lindquist coordinates $(t,r,\theta,\phi)$ takes the following form 
\begin{eqnarray}\label{Nmetric}
\mathrm{ds^2}=-\frac{\Delta_{r}}{\Sigma}\left(dt-\frac{a\sin^2\theta}{\Xi}d\phi\right)^2+\frac{\Sigma}{\Delta_{r}}dr^2+\frac{\Sigma}{\Delta_{\theta}}d\theta^2+\frac{\Delta_{\theta}\sin^2\theta}{\Sigma}\left(adt-\frac{a^2+r^2}{\Xi}d\phi\right)^2,
\end{eqnarray} 
with
\begin{equation}
\Sigma =r^2+a^2\cos^2\theta,\;\;
\Delta_{r}=\left(r^2+a^2\right)\left(1+\frac{r^2}{l^2}\right)-2Mre^{-k/r},
\label{deltar}
\end{equation}
and
\begin{equation}
\Delta_{\theta}=1-\frac{a^2}{l^2}\cos^2\theta,\;\;\Xi=1-\frac{a^2}{l^2}.
\end{equation}
The metric~(\ref{Nmetric}) depends on four parameters, namely the mass $M$, the spin $a$, the free parameter $k$ and the curvature radius $l$. The cosmological constant ($\Lambda$) is related to the radius of curvature $l$ via, $\Lambda=-3/l^2$. The metric~(\ref{Nmetric}) reduces to the Schwarzschild ($a=k=1/l^2=0$), the Schwarzschild-AdS ($a=k=0$), Kerr ($k=1/l^2=0$) and Kerr-AdS ($k=0$) black holes as special cases. The metric (\ref{Nmetric}) becomes singular apparently at $\Sigma=0$ and $\Delta_r=0$, but $\Delta_r=0$ is a coordinate singularity. It is seen that the metric (\ref{Nmetric}), for $k<<r$ reduces to the Kerr-Newman-AdS black hole such that \cite{Plebanski}
\begin{eqnarray}
\Delta_r=\left(r^2+a^2\right)\left(1+\frac{r^2}{l^2}\right)-2Mr+q^2+\mathcal{O}(k^2/r^2).
\end{eqnarray} 
The determinant of the metric (\ref{Nmetric}) takes the form
\begin{eqnarray}
\label{detg}
\sqrt{-g}=\frac{\Sigma\sin\theta}{\Xi}
\end{eqnarray}
Clearly, Eq.~\ref{Nmetric} becomes singular if $a^2=l^2$. At this critical condition the boundary of
AdS spacetime rotates at the speed of light. Therefore one must have to restrict $a^2<l^2$.
The stationary and axially symmetric spacetime (11) admits two commuting Killing vectors having the form $\xi^{\mu}{_(t)}=\delta^{\mu}{_t}$, and $\xi^{\mu}{_(\phi)}=\delta^{\mu}{_\phi}$ which leads to obtaining various metric components as a scalar product such that
\begin{eqnarray}
\label{Nmcomp}
\xi^{\mu}{_{(t)}}\xi_{(t){_\mu}}&=&g_{tt}=-1+\frac{2Mre^{-k/r}}{\Sigma}-\frac{r^2+a^2\sin^2\theta}{l^2},\\
\xi^{\mu}{_{(t)}}\xi_{(\phi){_\mu}}&=&g_{t\phi}=-\frac{a\sin^2\theta}{\Xi}\left(\frac{2Mre^{-k/r}}{\Sigma}-\frac{r^2+a^2}{l^2}\right),\\
\xi^{\mu}{_{(\phi)}}\xi_{(\phi){_\mu}}&=&g_{\phi\phi}=\frac{\sin^2\theta}{\Xi}\left(\left(r^2+a^2\right)\Xi+\frac{2Mra^2\sin^2\theta e^{-k/r}}{\Sigma}\right).
\end{eqnarray}
Using Einstein equation with a cosmological constant
\begin{equation}
G_{\mu\nu}=R_{\mu\nu}-\dfrac{1}{2}R g_{\mu\nu}+\Lambda g_{\mu\nu}=8\pi T_{\mu\nu},
\label{AdS3}
\end{equation}
We have the Einstein tensor as \cite{Xu:2016jod}
\begin{eqnarray}
G_{tt}&=&\dfrac{2[r^{4}-2r^{3}m+a^{2}r^{2}-a^{4}\sin^{2}\theta \cos^{2}\theta]m^{'}}{\Sigma^{3}}-\dfrac{ra^{2}\sin^{2}\theta m^{''}}{\Sigma^{2}}+\Lambda\dfrac{a^{2}\sin^{2}\theta-\Delta_{r}}{\Sigma},\nonumber\\
G_{rr}&=&-\dfrac{2r^{2}m^{'}}{\Sigma\Delta_{r}}+\Lambda\dfrac{\Sigma}{\Delta_{r}},~~~~G_{\theta\theta}=-\dfrac{2a^{2}\cos^{2}\theta m^{'}}{\Sigma}-rm^{''}+\Lambda\Sigma\nonumber,\\
G_{t\phi}&=&\dfrac{2a \sin^{2}\theta[(r^{2}+a^{2})(a^{2}\cos^{2}\theta-r^{2})]m^{'}}{\Sigma^{3}}-\dfrac{ra^{2}\sin^{2}\theta(r^{2}+a^{2})m^{''}}{\Sigma^{2}}+\Lambda\dfrac{a \sin^{2}\theta[\Delta_{r}-r^{2}-a^{2}]}{\Sigma},\nonumber\\
G_{\phi\phi}&=&-\dfrac{a^{2}\sin^{2}\theta[(r^{2}+a^{2})(a^{2}+(2r^{2}+a^{2})\cos2\theta)+2r^{3}\sin^{2}\theta m)]m^{'}}{\Sigma^{3}}-\dfrac{r\sin^{2}\theta(r^{2}+a^{2})^{2}m^{''}}{\Sigma^{2}}\nonumber\\
&&+\Lambda\dfrac{\sin^{2}\theta[(r^{2}+a^{2})^{2}-a^{2}\Delta_{r}]}{\Sigma}.
\end{eqnarray}

The two Killing vectors $\xi^{\mu}{_{(t)}}$ and $\xi^{\mu}{_{(\phi)}}$ denote the Killing symmetries corresponding to the time-translational symmetry along $t$-axis and the rotational symmetry around $\phi$-axis, together give rise to a Killing vector field $\chi^\mu=\xi^{\mu}{_{(t)}}+\Omega_+\xi^{\mu}{_{(\phi)}}$, which a null generator of the event horizon when $\Delta_r=0$, i.e., it is tangent to the null surface of the event horizon. The quantity $\Omega_+$ is the angular velocity at the event horizon of the black hole which reads as
\begin{eqnarray}
\label{omega1}
\Omega_+=\frac{a\Xi}{a^2+r_+^2}.
\end{eqnarray} 
One can also derive the angular velocity $\Omega$ of the observer moving on the orbits of constant $r$ and $\theta$ with four-velocity $u^{\mu}$ such that $u^{\mu}\xi_{(\phi){_\mu}}=0$ which turns out to be
\begin{eqnarray}
\label{omega2}
\Omega=-\frac{g_{t\phi}}{g_{\phi\phi}}=\frac{a\Xi\left(\left(r^2+a^2\right)\Delta_\theta-\Delta_r\right)}{\left(r^2+a^2\right)^2\Delta_\theta-a^2\sin^2\theta\Delta_r}.
\end{eqnarray}
In case of rotating regular AdS black hole the angular velocity (\ref{omega2}) does not vanish at asymptotic infinity rather it will give us a finite quantity so that
\begin{eqnarray}
\label{omega3}
\Omega_\infty=-\frac{a}{l^2}.
\end{eqnarray}
Now the Eqs.~(\ref{omega2}) and (\ref{omega3}) help us to define the angular velocity for the black holes with respect to a frame that is static at asymptotic infinity which gives
\begin{eqnarray}
\label{omega4}
\omega_+=\Omega_+-\Omega_\infty=\frac{a}{a^2+r_+^2}\left(1+\frac{r_+^2}{l^2}\right).
\end{eqnarray}
The angular velocity is the most important characteristic of the rotating
AdS black holes since it enters in the first law of thermodynamics to have a consistent relation. On the other hand, it can be shown that angular velocity coincides with that of the boundary Einstein universe, thereby providing the relevant basis for a AdS/CFT duality of rotating regular AdS black hole.

The metric (\ref{Nmetric}) belongs to the family of a non-Kerr black hole which encompasses Kerr as a special case ($k=1/l^2=0$). The non-Kerr black holes put some bounds on the spin parameter and other deviation parameters to behave like a Kerr black hole. The spacetime geometry around black hole candidates with the available radio and X-ray data \cite{Bambi:2013sha, Bambi:2013qj, Bambi:2011mj, Cardenas-Avendano:2016zml} shows us possible deviations from the black hole in GR. Proceeding along this line we consider the regular rotating AdS black hole candidates which may discard or constrain possible deviations from the Kerr background. 
\section{Null geodesics \label{sect3}}
In order to describe the black hole  shadow, we require the complete study of geodesic equations of photons around the black hole. The shadow images can provide the apparent shape of event horizon which is related to the geodesic structure of massless particle in the vicinity of the black hole. To study the geodesics of phpton   around rotating regular AdS black hole, we choose Hamilton-Jacobi equation due to Carter \cite{Carter:1968rr}. The complete geodesic equations of motion can be derived by solving  Hamilton-Jacobi equations
 \begin{eqnarray}
\label{NHmaJam}
\frac{\partial {\mathcal{S}}}{\partial \sigma} = -\frac{1}{2}g^{\alpha\beta}\frac{\partial {\mathcal{S}}}{\partial x^\alpha}\frac{\partial {\mathcal{S}}}{\partial x^\beta},
\end{eqnarray}
where $\sigma$ is the affine parameter alongside the geodesics and $g^{\alpha\beta}$ is the inverse metric tensor. The rotating regular AdS spacetime (\ref{Nmetric}) has two Killing vectors corresponding to two conserved quantities, energy $ E$ and the angular momentum $ L$. We choose the following separable solution  \cite{Chandrasekhar:1992}
\begin{eqnarray}
\mathcal{S}=\frac12 {m_0}^2 \sigma -{ E} t +{ L} \phi +\mathcal{S}_r(r)+\mathcal{S}_\theta(\theta) \label{Naction},
\end{eqnarray}
where $m_{0}$ is the mass of the test particle which is zero for photon case. Solving the Hamilton-Jacobi equation (\ref{NHmaJam}) using variable separable solution (\ref{Naction}), we derive the complete geodesic equations of motion
\begin{equation}
\Sigma\frac{dt}{d\sigma}=\frac{r^2+a^2}{\Delta_r}\left[E\left(r^2+a^2\right)-a\Xi L\right]-\frac{a}{\Delta_\theta}\left(aE\sin^2\theta-\Xi L\right),
\label{dott}
\end{equation}
\begin{equation}
\Sigma\frac{d\phi}{d\sigma}=\frac{a\Xi}{\Delta_r}\left[E\left(r^2+a^2\right)-a\Xi L\right]-\frac{\Xi}{\Delta_\theta}\left(aE-\Xi L\csc^2\theta\right),
\label{dotphi}
\end{equation}
\begin{equation}\label{NR}
\Sigma\frac{d r}{d\sigma}=\sqrt{R},
\end{equation}
and
\begin{equation}
\Sigma\frac{d\theta}{d\sigma}=\sqrt{\Theta},
\end{equation}
with
\begin{equation}
R=\left[(a^2 +r^2)E -a\Xi L\right]^2-\left[\left(L-aE\right)^2+\mathcal{K}\right]\Delta_r
\end{equation}
and
\begin{equation}
\Theta=-\left(a\sin^2\theta-\xi\Xi\right)^2\csc^2\theta-\left[\left(\xi-a\right)^2+\mathcal{K}\right]\Delta_{\theta},
\end{equation}
where $\mathcal{K}$ is the Carter separable constant. The above geodesic equations describes the motion of photons around the rotating regular AdS black hole. To study the photon orbits we introduce two impact parameters in term  of constants $ {E}$, $ {L}$ and $\mathcal{K}$ as $\xi= {L}/ {E}$ and $\eta=\mathcal{K}/{ {E}}^2$. The radial equation of motion (\ref{NR}) can be rewritten in terms of impact parameters $\eta$ and $\xi$ 
\begin{equation}\label{nrr}
R=\left[(a^2 +r^2)-a\xi\Xi\right]^2-\left[\left(\xi-a\right)^2+\eta\right]\Delta_r.
\end{equation} 
The apparent shape of black hole shadow can be determined by some constant values  of the orbits $r=r_{p}$, which must satisfy the following conditions
\begin{equation}
\label{vr}
 \mathcal{R}(r)=\frac{\partial \mathcal{R}(r)}{\partial r}=0,
\end{equation} 
solving these equations simultaneously one can get the impact parameters $\eta$ and $\xi$ 
\begin{eqnarray}
\xi(r_p)&=& \frac{-4 r_p \Delta_r+\left(r_p^2+a^2\right)\Delta_r{^\prime}}{a \Xi  \Delta_r{^\prime}},\label{Nxi}  \\
 \eta(r_p)&=&\frac{-16 r_p^2 \Delta_r^2-\left(r_p^2+\frac{a^4}{l^2}\right)^2 \Delta_r{^\prime}^2+8 r_p \Delta_r \left(2 a^2 r_p \Xi ^2+\left(r_p^2+\frac{a^4}{l^2}\right)\Delta_r{^\prime}\right)}{a^2 \Xi ^2 \Delta_r{^\prime}^2}, \nonumber \label{Neta}
 \\
\end{eqnarray} 
where the term $\Delta_r{^\prime}$ denotes the derivative of $\Delta_r$ with respect to $r$ at point $r=r_p$ and it can be written as 
\begin{eqnarray}
\Delta_r{^\prime}=-\frac{2}{3}\left(3\left(Me^{-k/r_p}(k+r_p)/r_p\right)-3r_p-\frac{1}{3l^2} r_p(a^2+2r_p^2)\right).
\end{eqnarray}
The geometry of photon orbits around regular AdS black hole can be determined by the impact parameters given in Eqs.~(\ref{Nxi}) and (\ref{Neta}). In the absence of cosmological constant, these impact parameters reduce to the expressions for nonsingular spacetime \cite{Amir:2016cen} as 
\begin{eqnarray}
\xi &=& \frac{3 M \left(a^2 (k+r_p)+r_p^2 (k-3 r_p)\right)+3 r_p^2 \left(a^2+r_p^2\right) e^{k/r_p}}{3 a M (k+r_p)-3 a r_p^2 e^{k/r_p}}, \label{xiexp}\\
\eta &=&-\frac{r_p^4 \left(4 a^2 M e^{k/r_p} (k-r_p)+\left(r_p \left(r_p e^{k/r_p}-3 M\right)+k M\right)^2\right)}{a^2 \left(r_p \left(M-r_p e^{k/r_p}\right)+k M\right)^2}, 
\end{eqnarray}
and which in addition, for $k=0$ exactly reduces to the Kerr spacetime such that
 \begin{eqnarray}
\xi &=& \frac{a^2 (M+r_p)+r_p^2 (r_p-3 M)}{a (M-r_p)}, \label{nkxi}\\
\eta &=&-\frac{r_p^3 \left(r_p (r_p-3 M)^2-4 a^2 M\right)}{a^2 (M-r_p)^2} \label{nketa}.
\end{eqnarray}
The expressions of $\eta$ and $\xi$ in Eqs~(\ref{Nxi}) and (\ref{Neta}) completely describe the photon orbits. In order to determine the shadow images observed on the sky at some finite distant one has to use the coordinates, say ($r_O,\;\vartheta_O$), related to the orthonormal tetrads.
\begin{figure*}
    \begin{tabular}{c c c c}
	\includegraphics[scale=0.51]{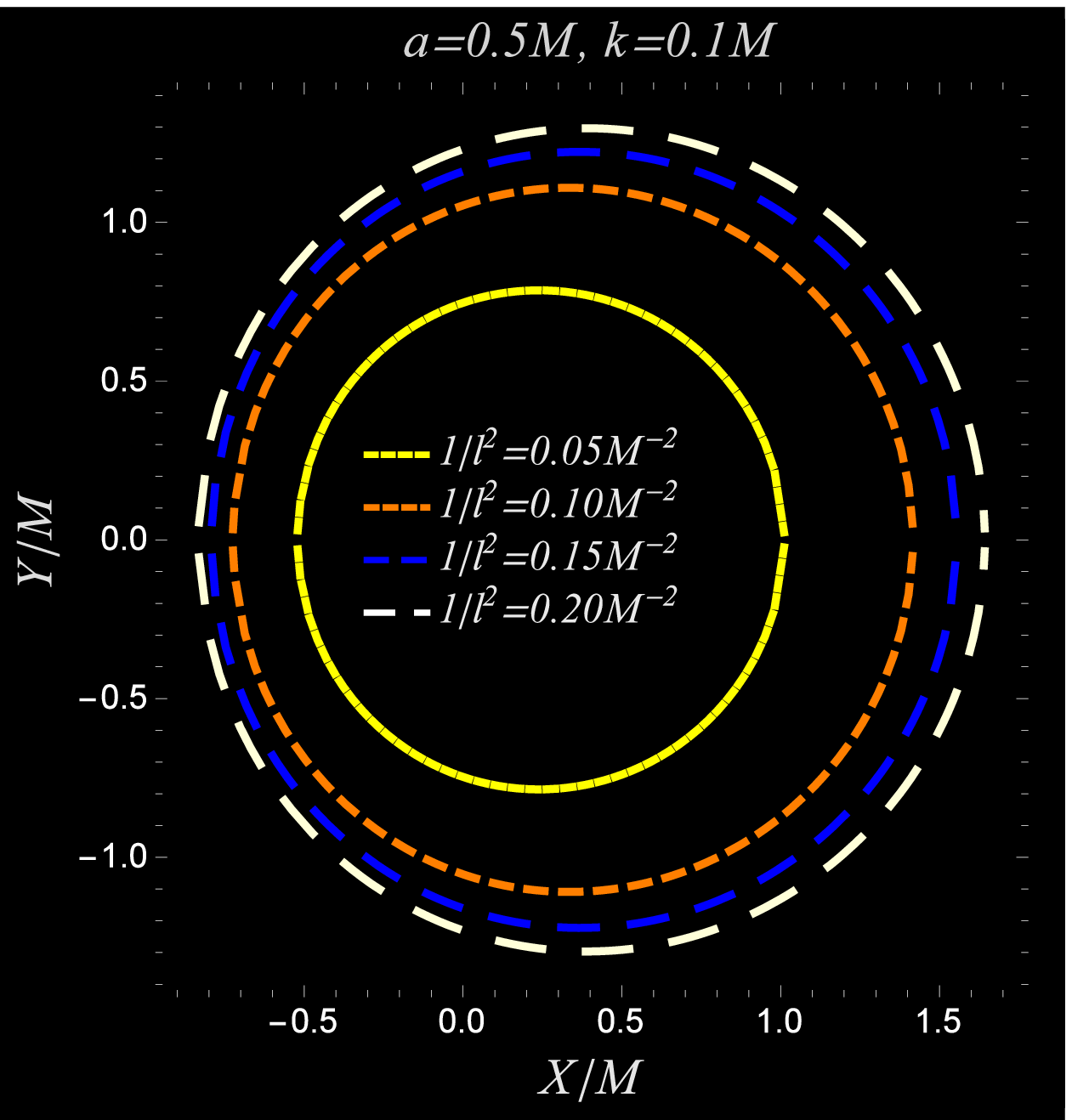} 
	\includegraphics[scale=0.5]{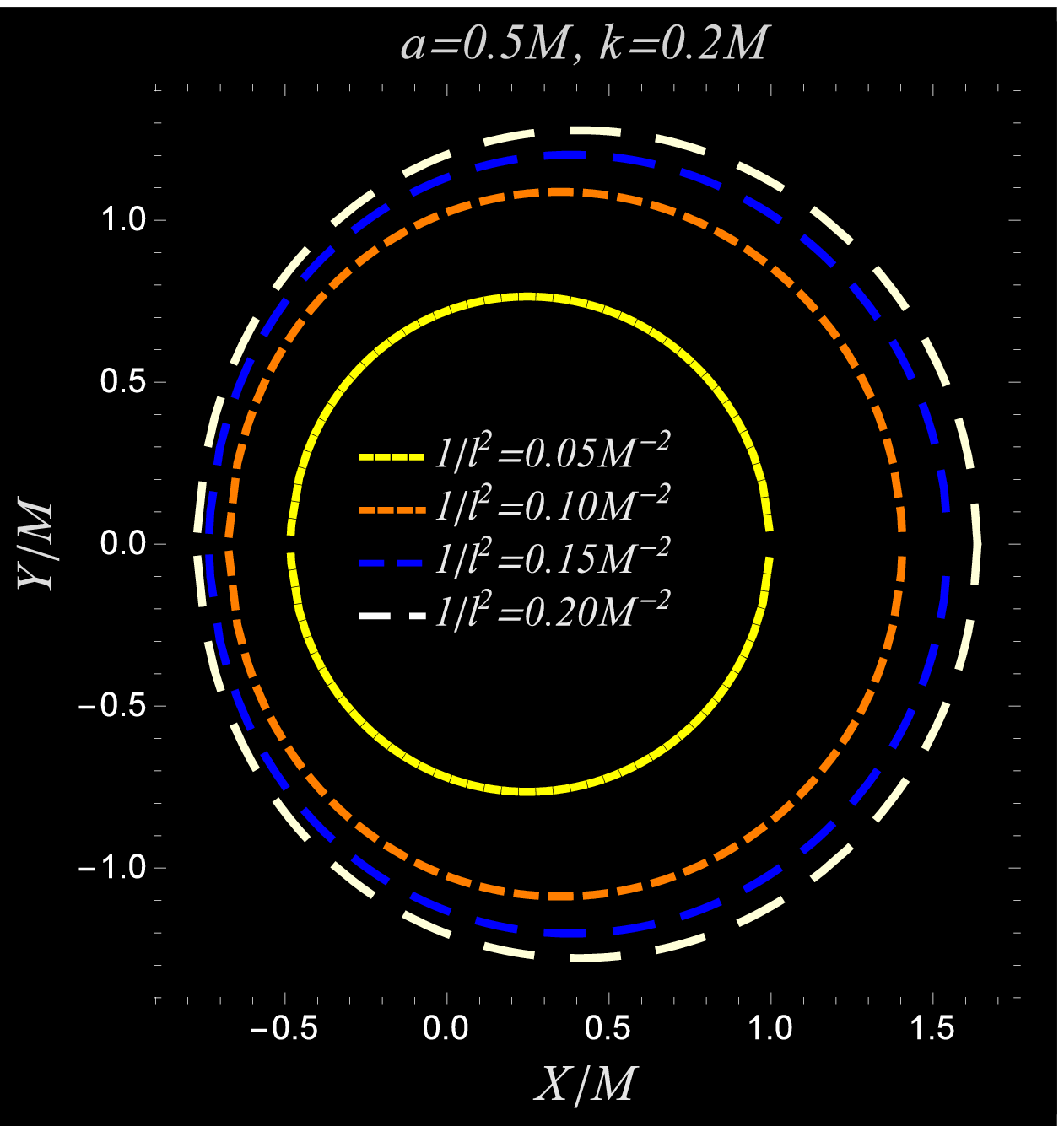} \\
     \includegraphics[scale=0.51]{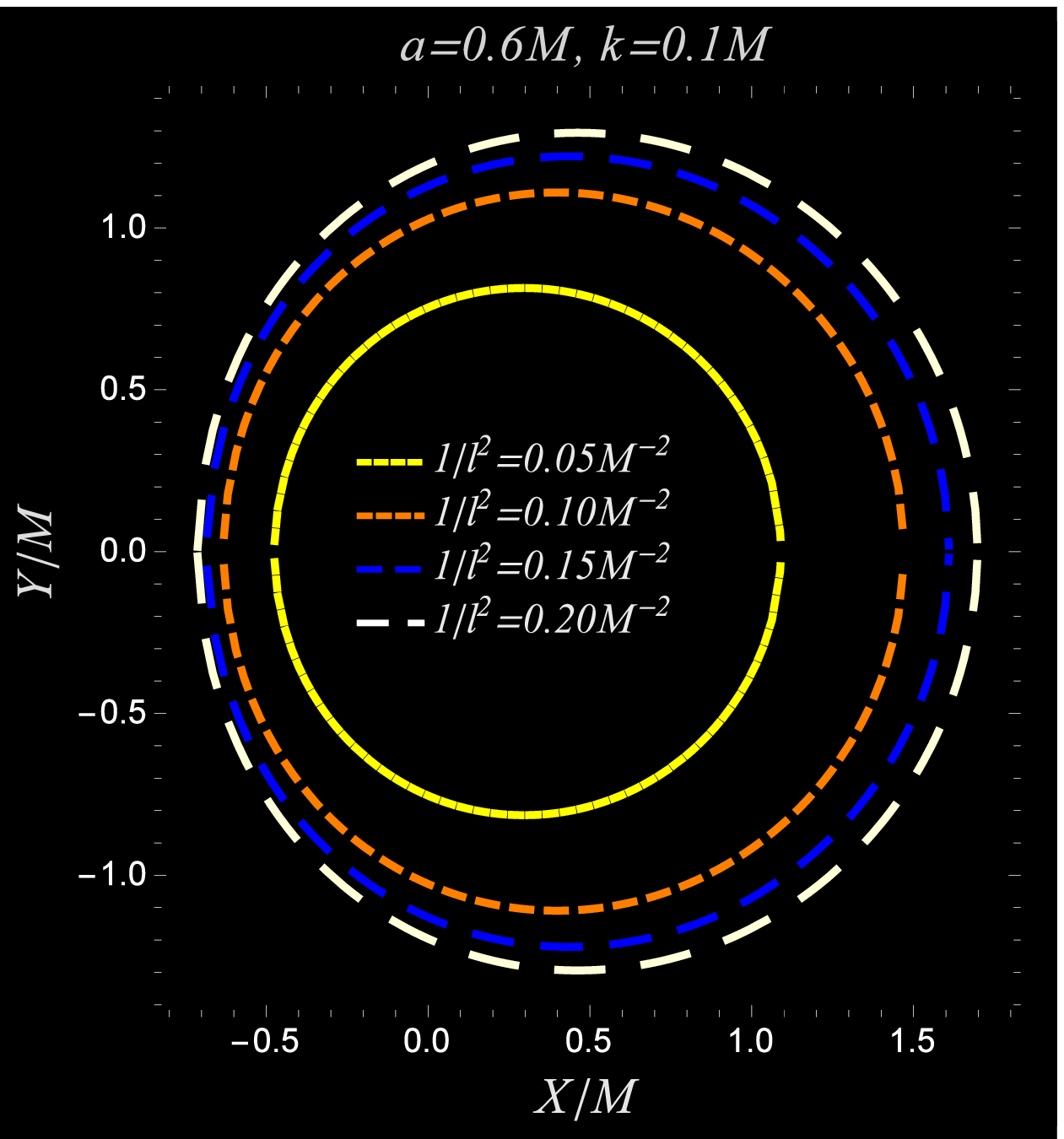}
	\includegraphics[scale=0.5]{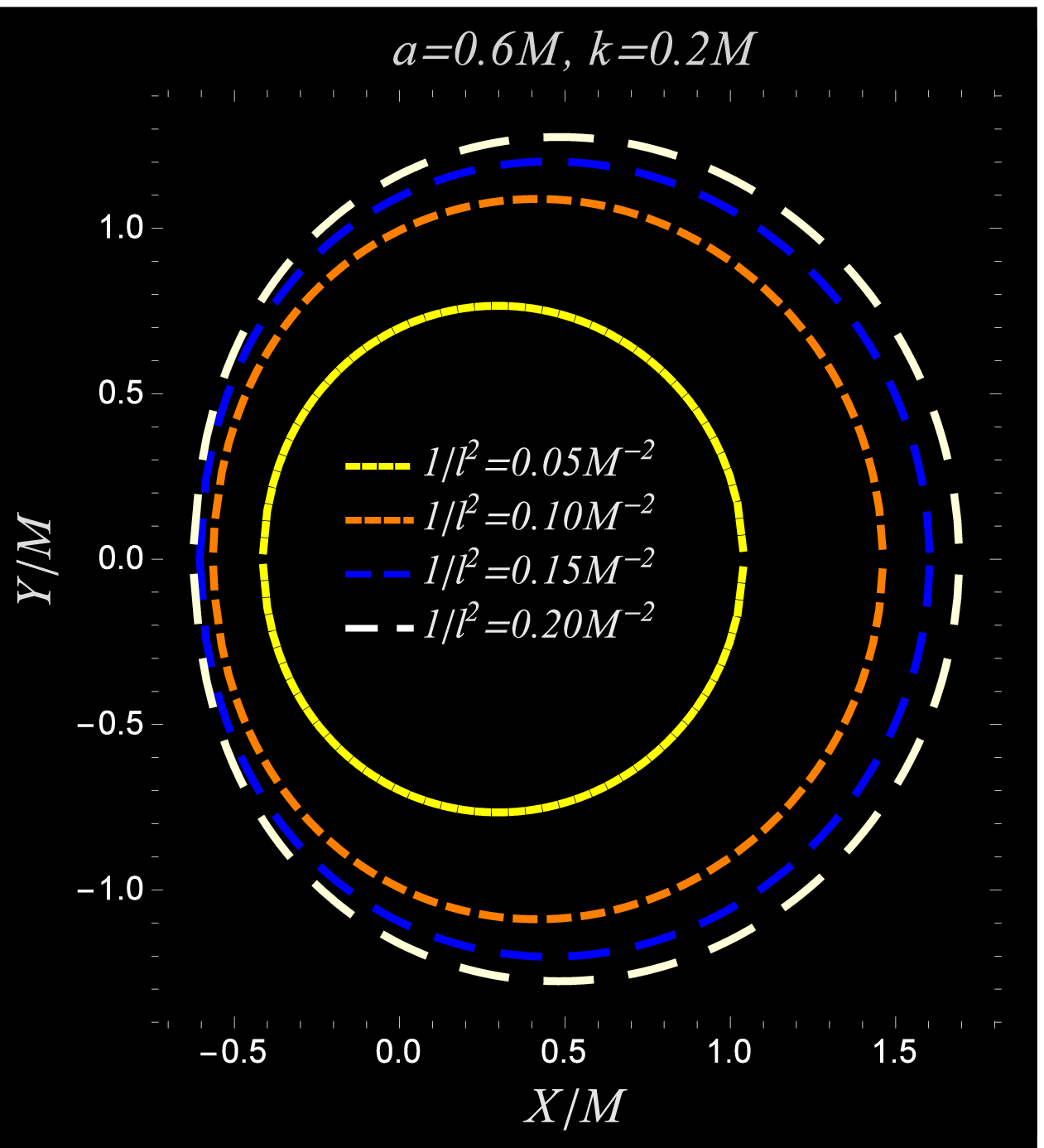} 
	 \end{tabular}
   \caption{\label{Nfig2}   Plot showing the black hole shadow in rotating regular AdS spacetime at $r_O=50$ and for different values of $k$, $a$ and $1/l^2$.}
\end{figure*}

\begin{figure*}
    \begin{tabular}{c c c c}
	\includegraphics[scale=0.51]{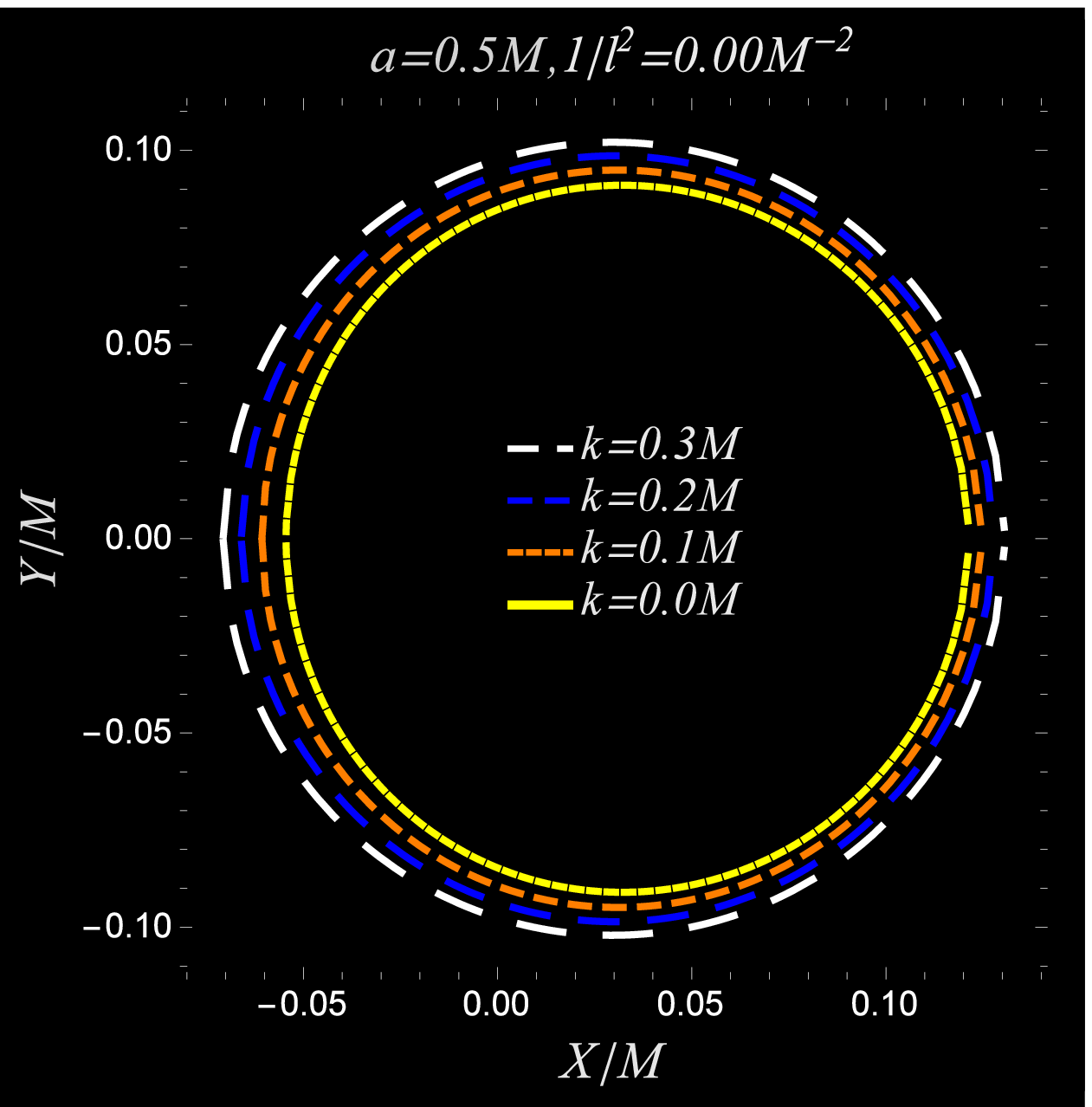} 
	\includegraphics[scale=0.51]{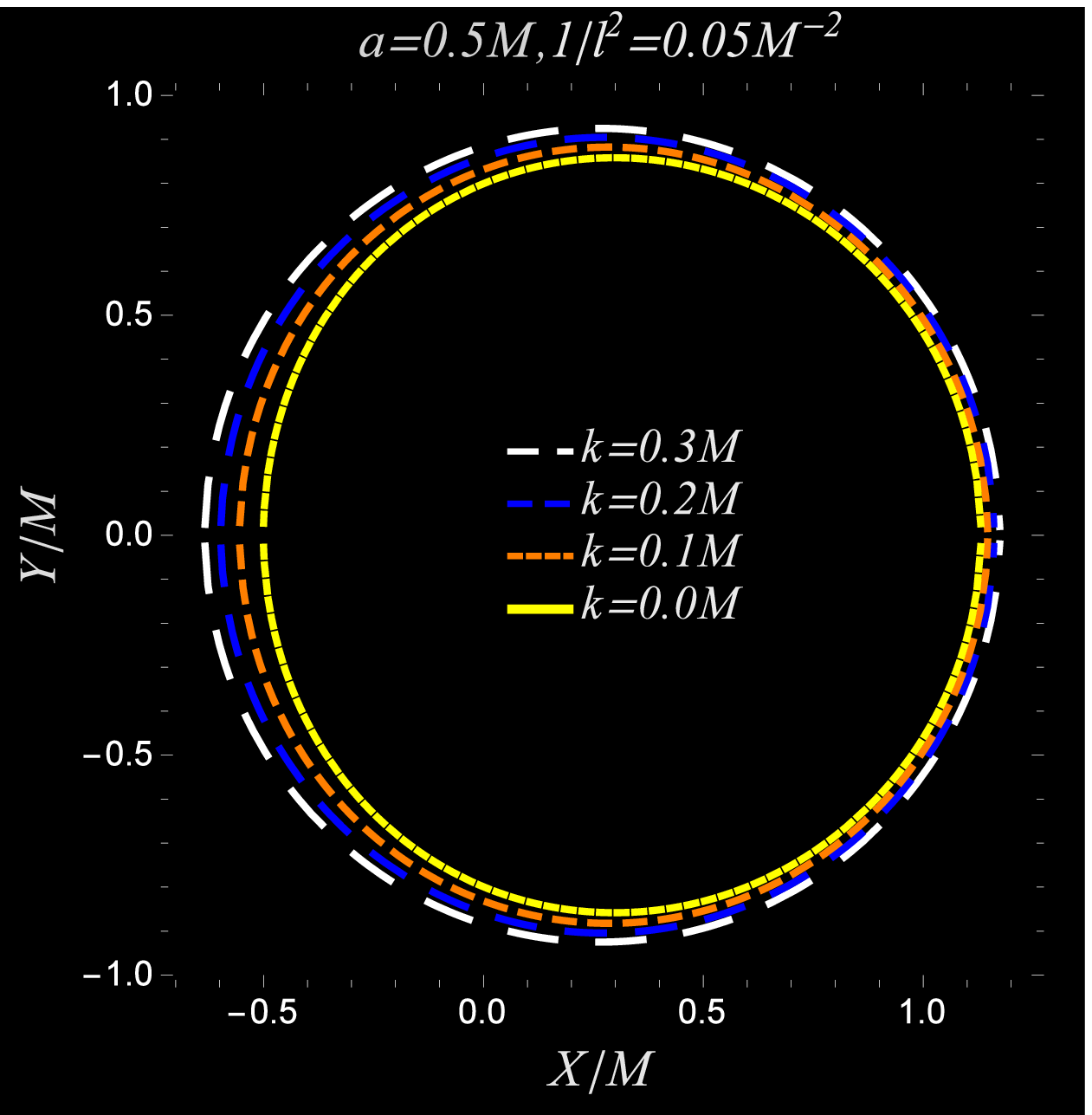} \\
	\includegraphics[scale=0.51]{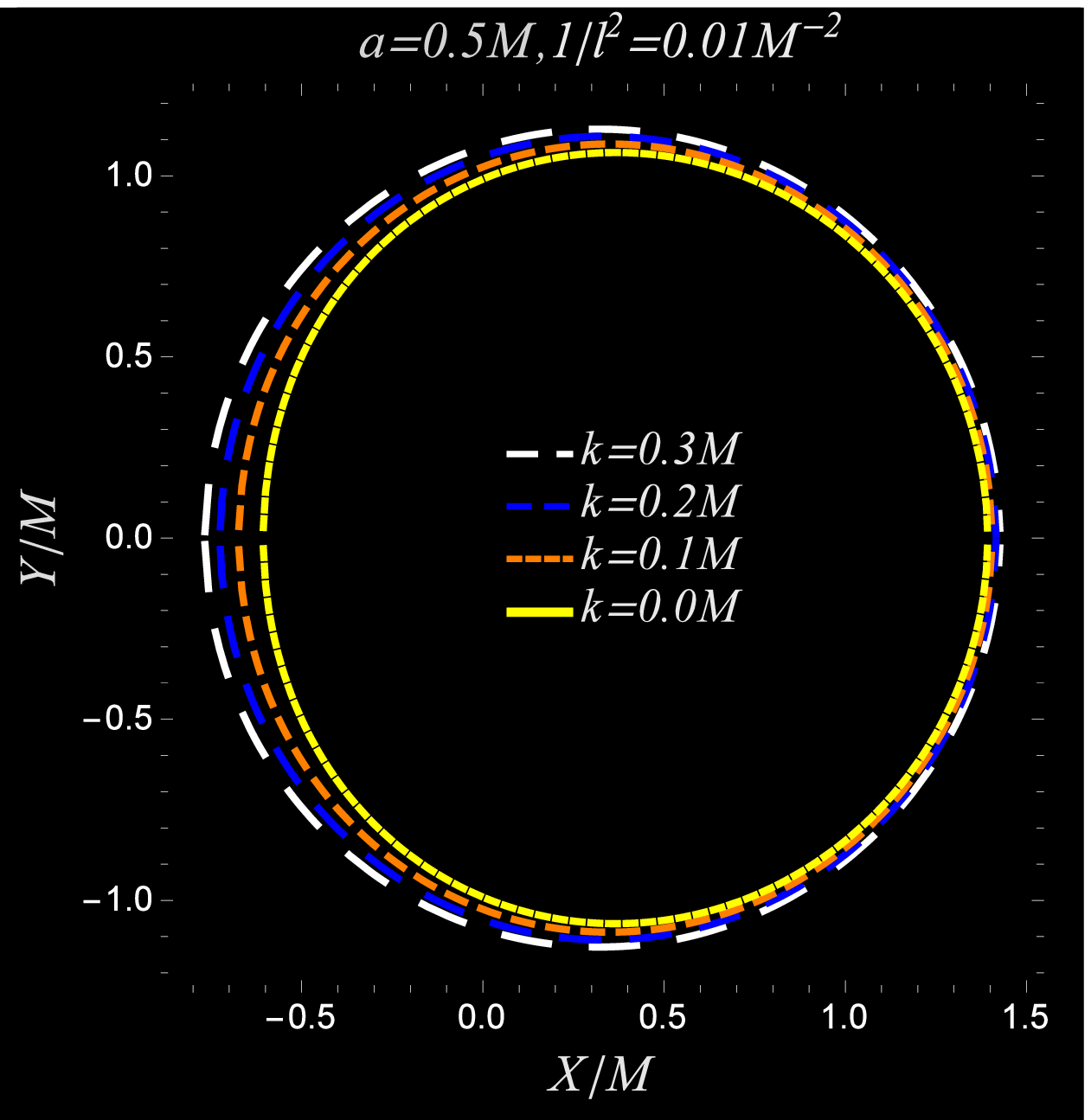} 
	\includegraphics[scale=0.51]{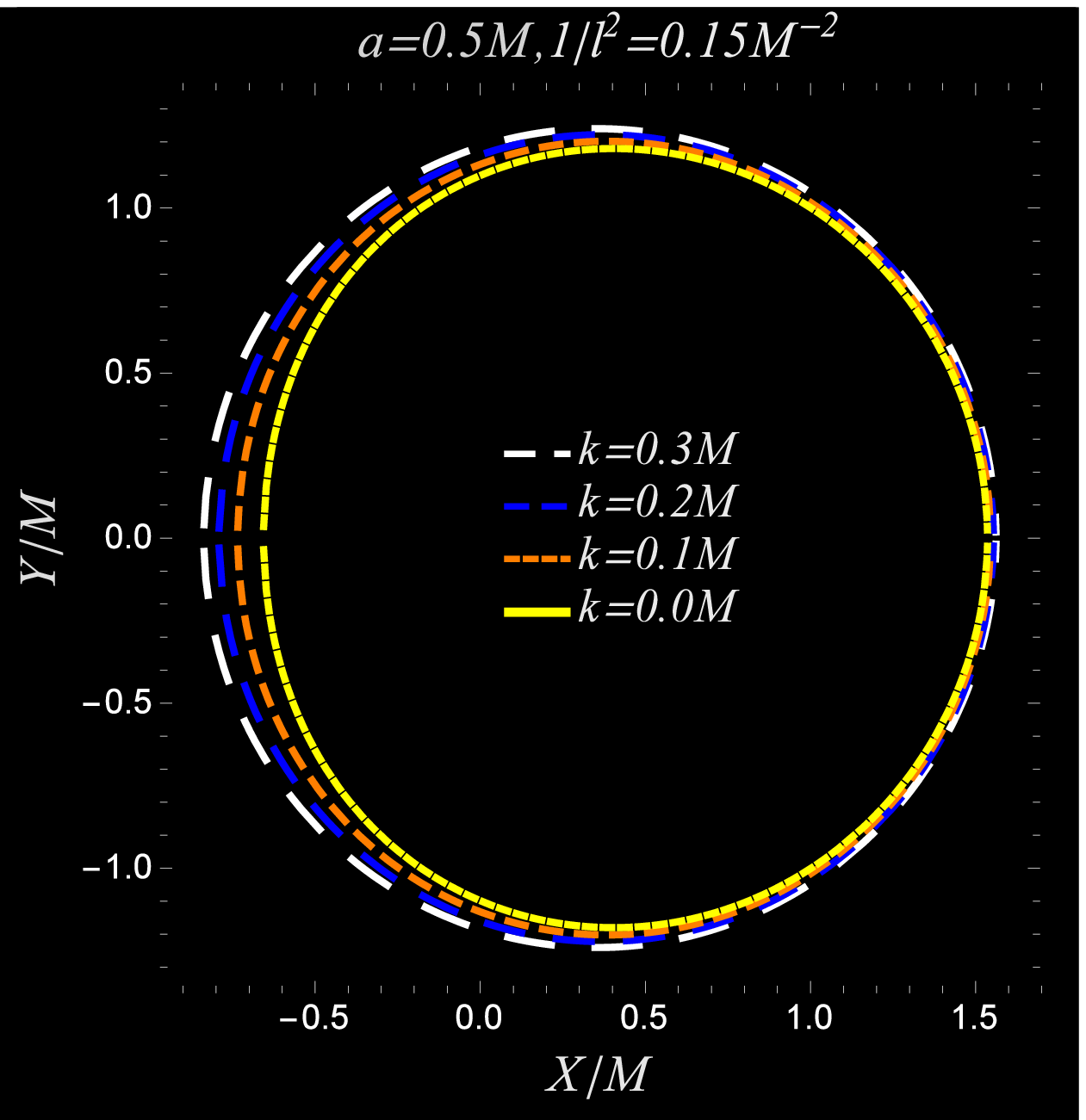} 
    \end{tabular}
        \caption{\label{Nfig3}   Plot showing the black hole shadow in rotating regular AdS spacetime at $r_O=50$ and for different values of $k$, $a$, and $1/l^2$.}
\end{figure*}
\section{Rotating regular AdS black hole shadow \label{sect4}}
In this section, we construct the shadow of the rotating regular AdS black holes. For this, we fix the observer position at some given Boyer-Lindquist coordinates $(r_O,\vartheta_O)$ in the domain of outer communication. We define the orthonormal tetrads at $(r_O,\vartheta_O)$ as \cite{Grenzebach:2014fha, Eiroa:2017uuq}
\begin{equation}
e_0=\left.\frac{\left(r^2+a^2\right)\partial_t+a\Xi \partial_\phi}{\sqrt{\Delta_r\Sigma}}\right|_{(r_O,\vartheta_O)},\;\;e_1=\left.\sqrt{\frac{\Delta_\theta}{\Sigma}}\partial_\theta\right|_{(r_O,\vartheta_O)},
\label{e0}
\end{equation}
and
\begin{equation}
e_2=-\left.\frac{\Xi \partial_\phi+a\sin^2\theta \partial_t}{\sqrt{\Delta_\theta\Sigma}\sin\theta}\right|_{(r_O,\vartheta_O)},\;\; e_3=-\left.\sqrt{\frac{\Delta_r}{\Sigma}}\partial_r\right|_{(r_O,\vartheta_O)},
\end{equation}
where $e_0$ and  $e_3$  denotes the four velocity of the observer which gives the spatial direction towards the  black hole and the combination $e_0\pm e_3$ is tangential to the outgoing and ingoing principal null congruences of the metric. The coordinates $\lambda(s)=(r(s),\theta(s),\phi(s),t(s))$ be assigned for each light ray which in turn gives its tangent vector as
\begin{equation}
\dot{\lambda}=\dot{r}\partial_r+\dot{\theta}\partial_\theta+\dot{\phi}\partial_\phi+\dot{t}\partial_t,
\label{lambda1}
\end{equation}
and the tangent vector  in terms of the coordinates $\alpha$ and $\beta$ results
\begin{equation}
\dot{\lambda}=\gamma\left(-e_0+\sin\alpha\cos\beta e_1+\sin\alpha\sin\beta e_2+\cos\alpha e_3\right).
\label{lambda2}
\end{equation}
For each light ray the scalar factor $\gamma$ can be calculated from Eqs.~(\ref{lambda1}) and (\ref{lambda2}) as
\begin{equation}
\gamma=g(\dot{\lambda},e_0)=g\left(\dot{\lambda},\frac{r^2+a^2}{\sqrt{\Delta_r\Sigma}}\partial_t+\frac{a\Xi}{\sqrt{\Delta_r\Sigma}}\partial_\phi\right)=\left.\frac{-E\left(r^2+a^2\right)+a\Xi L}{\sqrt{\Delta_r\Sigma}}\right|_{(r_O,\vartheta_O)},
\end{equation}
and by comparing the corresponding coefficients in Eqs.~(\ref{lambda1}) and (\ref{lambda2}), we have
\begin{equation}
\sin\alpha=\left.\sqrt{1-\left[\frac{\dot{r}\Sigma}{E\left(r^2+a^2\right)-a\Xi L}\right]^2}\right|_{(r_O,\vartheta_O)}
\end{equation}
and
\begin{equation}
\sin\beta=\left.\frac{\sqrt{\Delta_\theta}\sin\theta}{\Xi\sqrt{\Delta_r}\sin\alpha}\left[\frac{\dot{\phi}\Sigma\Delta_r}{E\left(r^2+a^2\right)-a\Xi L}-a\Xi\right]\right|_{(r_O,\vartheta_O)},
\end{equation}
 by substituting the impact parameters (\ref{Nxi}) and (\ref{Neta}) in Eqs. (\ref{dotphi}) and (\ref{NR}), we find
\begin{equation}
\sin\alpha=\left.\frac{\sqrt{\left[\left(\xi(r_p)-a\right)^2+\eta(r_p)\right]\Delta_r}}{r^2+a^2-a\Xi \xi(r_p)}\right|_{(r_O,\vartheta_O)},
\end{equation}
and
\begin{equation}
\sin\beta=\left.\frac{\sqrt{\Delta_r}\sin\theta}{\sqrt{\Delta_\theta}\sin\alpha}\left[\frac{a-\Xi\xi(r_p)\csc^2\theta}{a\Xi\xi(r_p)-\left(r^2+a^2\right)}\right]\right|_{(r_O,\vartheta_O)}.
\end{equation}
 \begin{figure*}
    \begin{tabular}{c c c c}
	\includegraphics[scale=0.6]{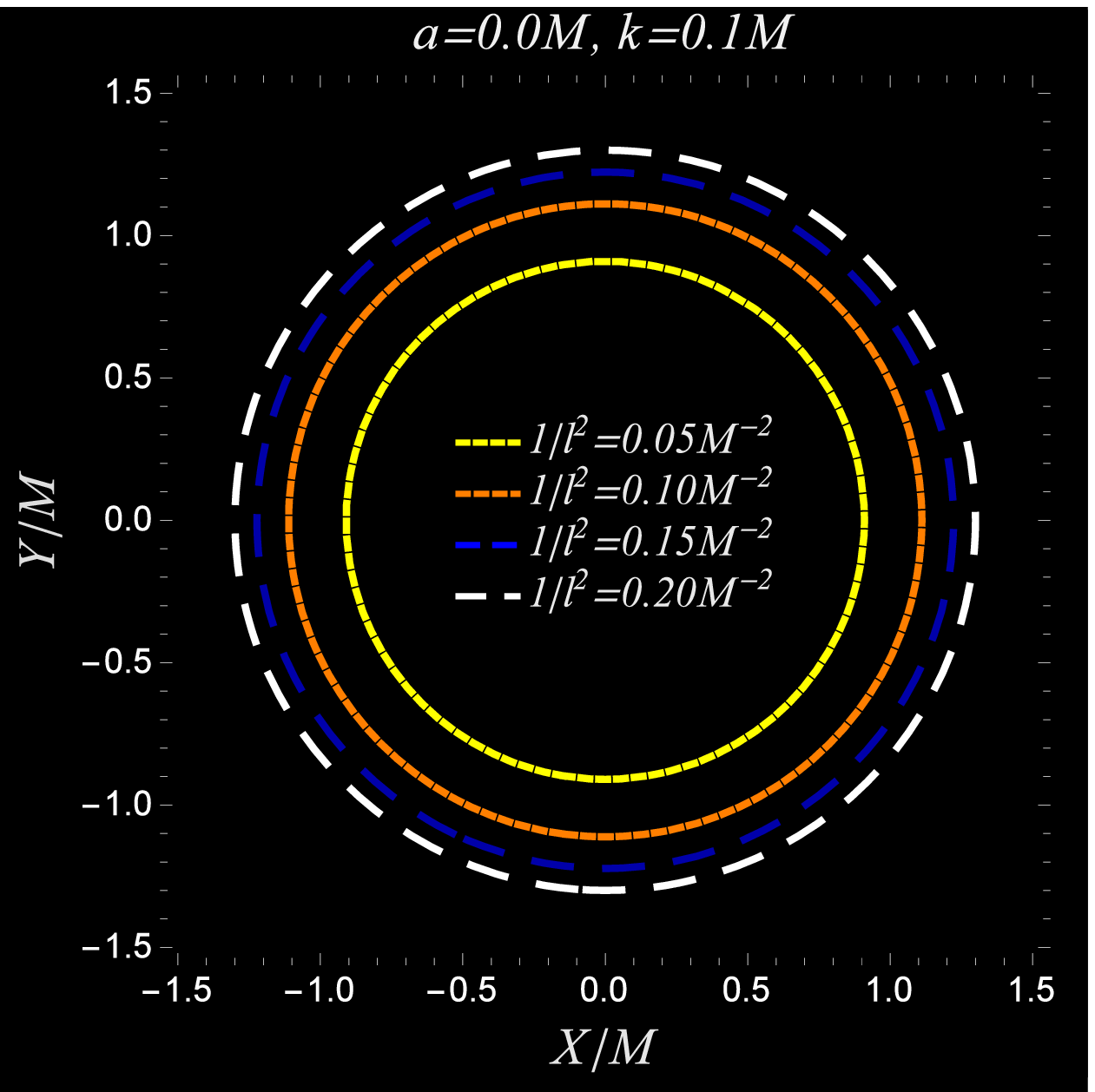} 
	\includegraphics[scale=0.6]{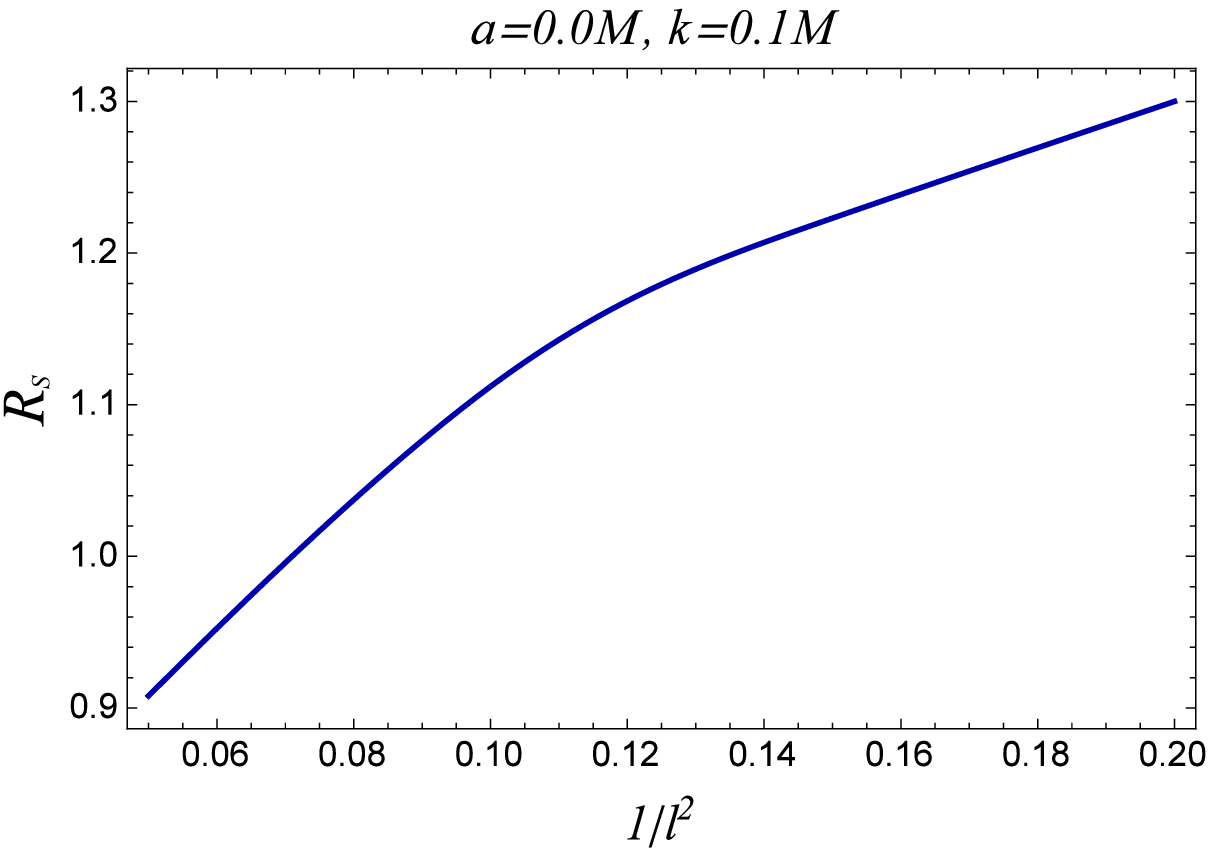} 
   	\end{tabular}
     \caption{\label{Nfig6} Plots showing the non-rotating regular AdS black hole shadow and its observable $R_s$   for different values of  $k$ with $1/l^2$ at $r_O=50$.}
\end{figure*}
In order to obtain the apparent shape of the shadow, we have considered the stereographic projection of celestial sphere onto a plane, where the Cartesian coordinates in this plane takes the form
\begin{eqnarray}
X=-2\tan\left(\frac{\alpha(r_p)}{2}\right)\sin\left(\beta(r_p)\right),\nonumber\\
Y=-2\tan\left(\frac{\alpha(r_p)}{2}\right)\cos\left(\beta(r_p)\right),
\label{ncount}
\end{eqnarray}
which satisfies the relation
\begin{eqnarray}
X^2+Y^2=4\tan^2\left(\frac{\alpha(r_p)}{2}\right).
\label{ncount}
\end{eqnarray}
\begin{figure*}
   \begin{tabular}{c c c c}
	\includegraphics[scale=0.55]{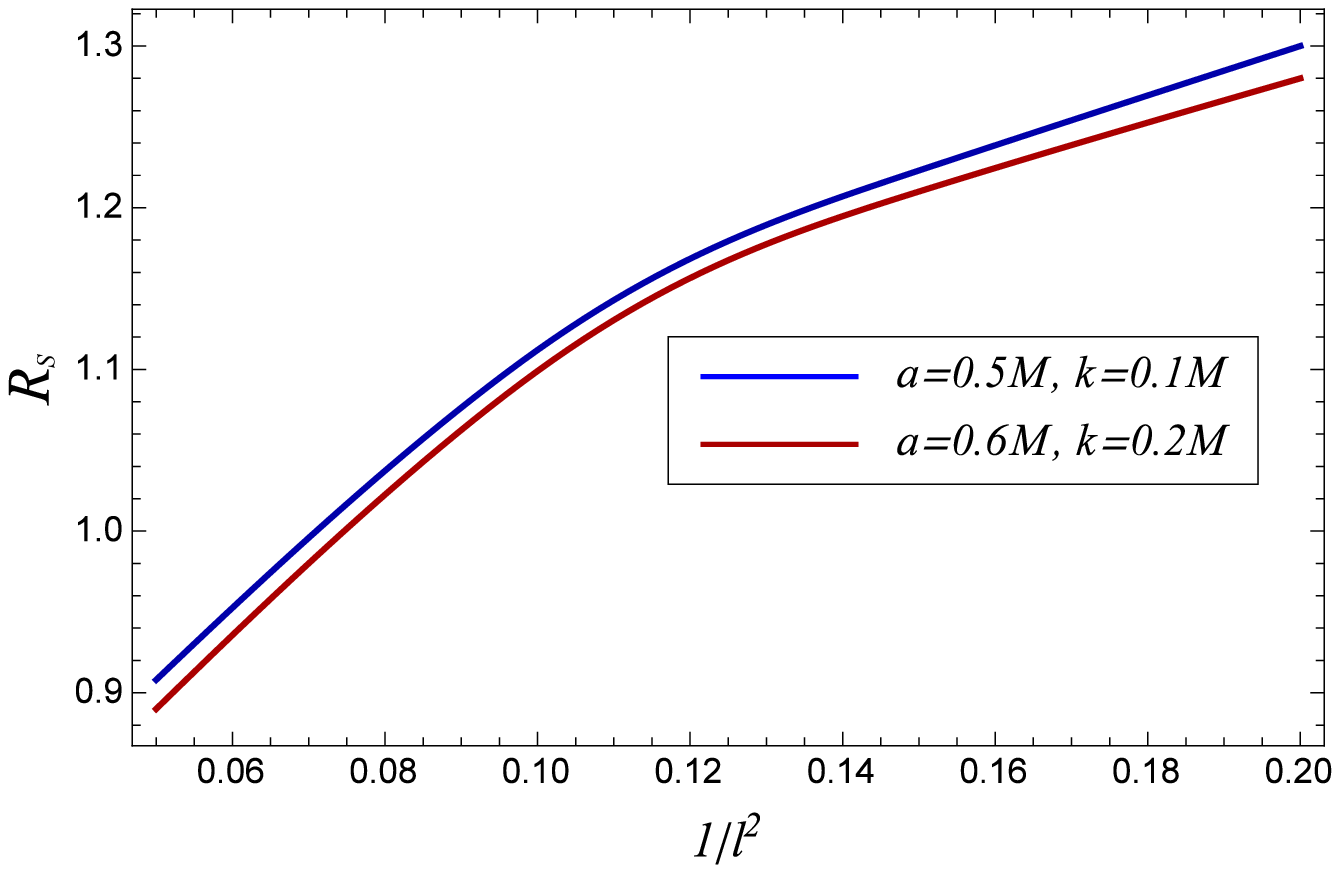} 
	\includegraphics[scale=0.55]{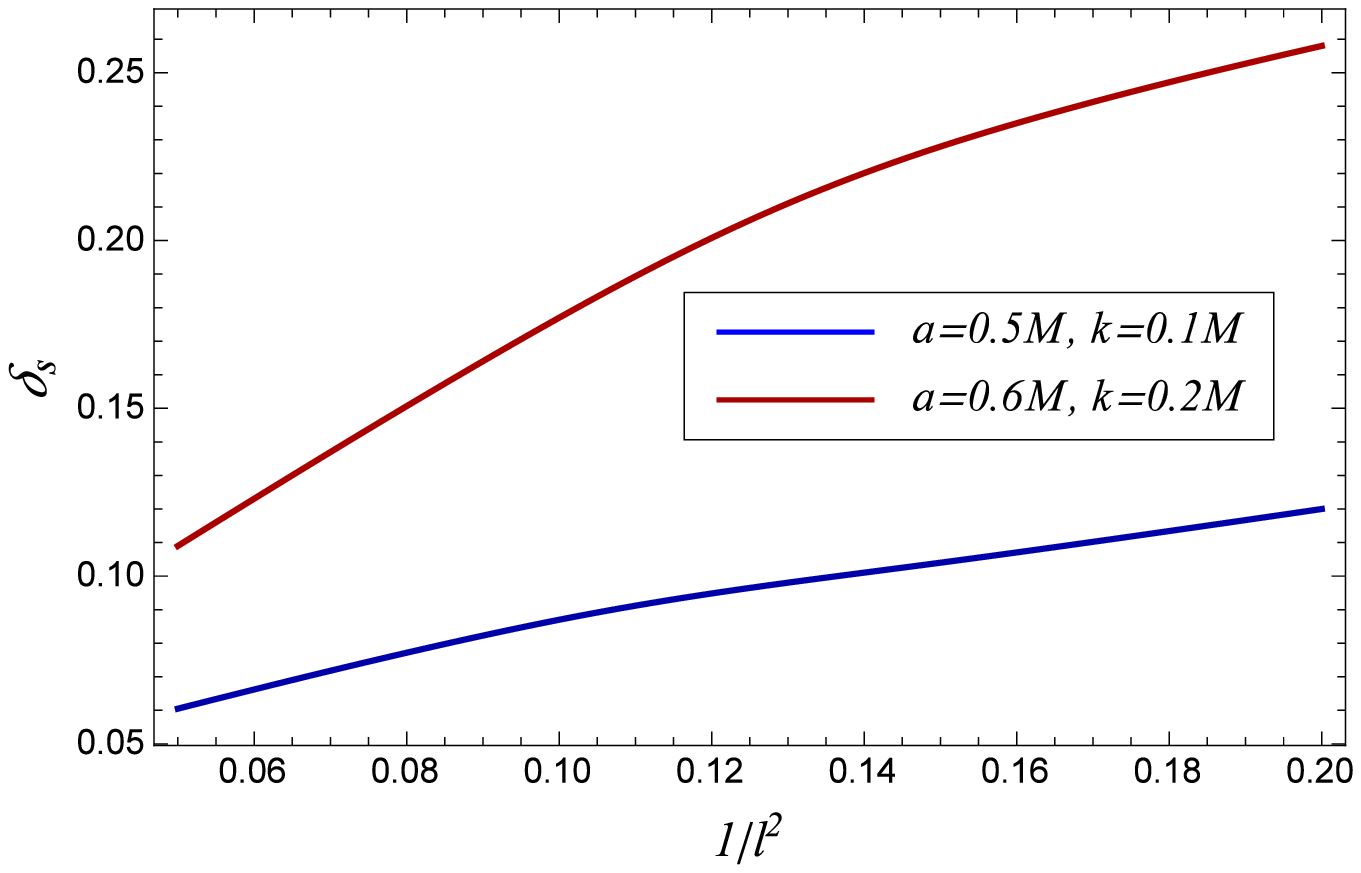} 
   	\end{tabular}
     \caption{\label{fig4} Plots showing the  variation of  observables $R_s$ and $\delta_s$ with $1/l^2$ for different values of $a$ and $k$ at $r_O=50$.}
\end{figure*}
\begin{figure*}
   \begin{tabular}{c c c c}
	\includegraphics[scale=0.65]{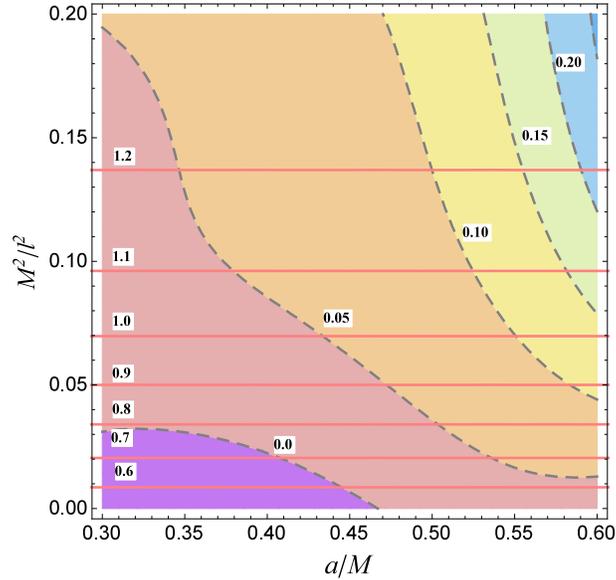} 
	 \end{tabular}
    \caption{\label{fig5}The contours of $R_s$ (solid red lines) and $\delta_s$ (dashed gray lines) in ($a, 1/l^2$) plane. The point of intersection of $R_s$ and $\delta_s$ lines gives the values of black hole parameters.}
\end{figure*}
The shapes of regular AdS black hole shadow can be found by plotting the contour of Eq.~(\ref{ncount}). The shape and size of the black hole shadow mainly depend on its spin parameter $a$, the curvature radius $1/l^2$ and free parameter $k$. We plot the contour of regular AdS black hole shadow in Fig.~(\ref{Nfig2}) and (\ref{Nfig3}) for various values of $a$, $1/l^2$, and $k$. The size of the black hole shadow and distortion increases for the increasing values of the $1/l^2$ (cf. Fig.~\ref{Nfig2}) while the size of black hole shadow decreases for the increasing values of free parameter $k$ (cf. Fig.~\ref{Nfig3}). We also plot the spherical ($a=0$) regular AdS black hole shadow in Fig.~(\ref{Nfig6}) which appears as a perfect circle and increases with the increasing values    $1/l^2$. 

Next, we define two observables, namely the shadow radius $R_s$ and the distortion parameter $\delta_s$  \cite{Hioki:2009na}. The observables $R_s$, which measures, can be defined \cite{Hioki:2009na}
\begin{equation}
R_s=\frac{(X_t-X_r)^2+{Y_t}^2}{2|{X_t}-{X_r}|},
\end{equation}
where $(X_t,\;Y_t)$ and $(X_r,\;Y_r)$ are the topmost and the rightmost positions of the celestial coordinates from where the contour of the black hole shadow passes. The second observable $\delta_s$ depend on the black hole shadow radius $R_s$ and the dent $D_s$ which occur due to the rotation of the black hole. The expression of $\delta_s$ reads 
\begin{equation}
\delta_s=\frac{D_s}{R_s}.
\end{equation} 
 \begin{center}
 \begin{table}  
 \begin{tabular}{|p{3cm}|p{3cm}|p{3cm}|}
 \hline
  \multicolumn{3}{|c|}{$ a=0.5M, k=0.1M $} \\
 \hline
 radius of curvature $(M^2/l^2)$& shadow radius $(R_s/M)$ & distortion parameter $(\delta_{s})$ \\
 \hline
\quad\quad0.05    & \quad\quad     0.90     &         \quad\quad     0.06   \\
 \quad\quad0.10   &  \quad\quad      1.11  &        \quad\quad     0.08   \\
\quad\quad0.15    &\quad\quad         1.22   &       \quad\quad      0.10 \\
 \quad\quad0.20    &\quad\quad      1.30    &       \quad\quad       0.12   \\\hline
\end{tabular}
\begin{tabular}{|p{3cm}|p{3cm}|p{3cm}|}
 \hline
  \multicolumn{3}{|c|}{$a=0.6M, k=0
  .2M$} \\
 \hline
radius of curvature $M^2/l^2$& shadow radius $(R_s/M)$ & distortion parameter $(\delta_{s})$ \\
 \hline
\quad\quad 0.05  & \quad\quad 0.89    &              \quad\quad 0.11   \\
 \quad\quad0.10&  \quad\quad 1.09      &           \quad\quad  0.17   \\
\quad\quad0.15 &\quad\quad1.21        &                  \quad\quad     0.23 \\
\quad\quad 0.20    &\quad\quad1.28       &        \quad\quad       0.27   \\
\hline
\end{tabular} 
\caption{Variation of shadow radius and distortion parameter with parameter $M^2/l^2$.  } \label{table1}  
 \end{table}
 \end{center}
In Fig.~\ref{fig4} we plot the variation of the black hole shadow radius $R_s$ and the distortion parameter $\delta_s$. For  fixed values of the spin parameter and free parameter, the radius of the  black hole shadow $R_s$ and the distortion parameter $\delta_s$  increases with the   radius of curvature parameter $1/l^2$ (cf.~Fig.~\ref{fig4}). In Fig.~{\ref{fig5}} we plot the shadow radius $R_s$ and distortion parameter $\delta_s$ in $(a,1/l^2)$ plane. The point of intersection of $R_s$ and $\delta_s$ lines gives the spin parameter $a$ and radius of curvature parameter $1/l^2$. If we compare our results   with nonsigular black hole shadow \cite{Amir:2016cen},  we find  in ads spacetime the size of black hole shadow appears larger and gets  more distorted for the increasing values of parameter $1/l^2$ (cf. Fig.~\ref{Nfig2}). In comparison with the Kerr black hole \cite{Chandrasekhar:1992}, our results shows that the size of shadow appears bigger in regular ads spacetime. A detailed numerical values of shadow radius and distortion parameter for different values of spin parameter, free parameter and  have shown in the Table~\ref{table1}.
\section{Energy emission rate}\label{sect5}
In the previous discussion, we derived the required derivation for the nonsingular black hole silhouettes in asymptotically anti-de sitter spacetimes. The silhouettes of the black holes are responsible for the high energy absorption cross-section for the asymptotically flat or (Anti)-de Sitter observers. The absorption cross section of the black holes slightly modulates near a limiting constant value at high energies. 
 \begin{figure*}
  \begin{tabular}{c c c c}
	\includegraphics[scale=0.65]{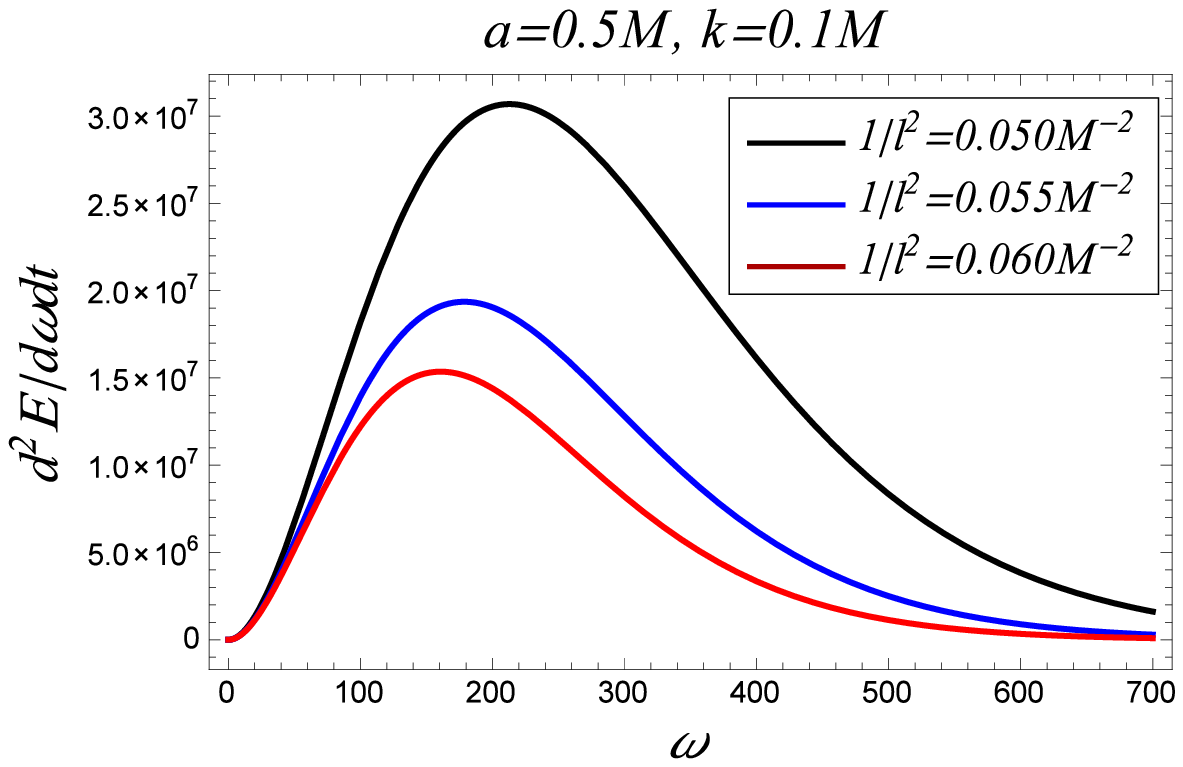} 
	\includegraphics[scale=0.65]{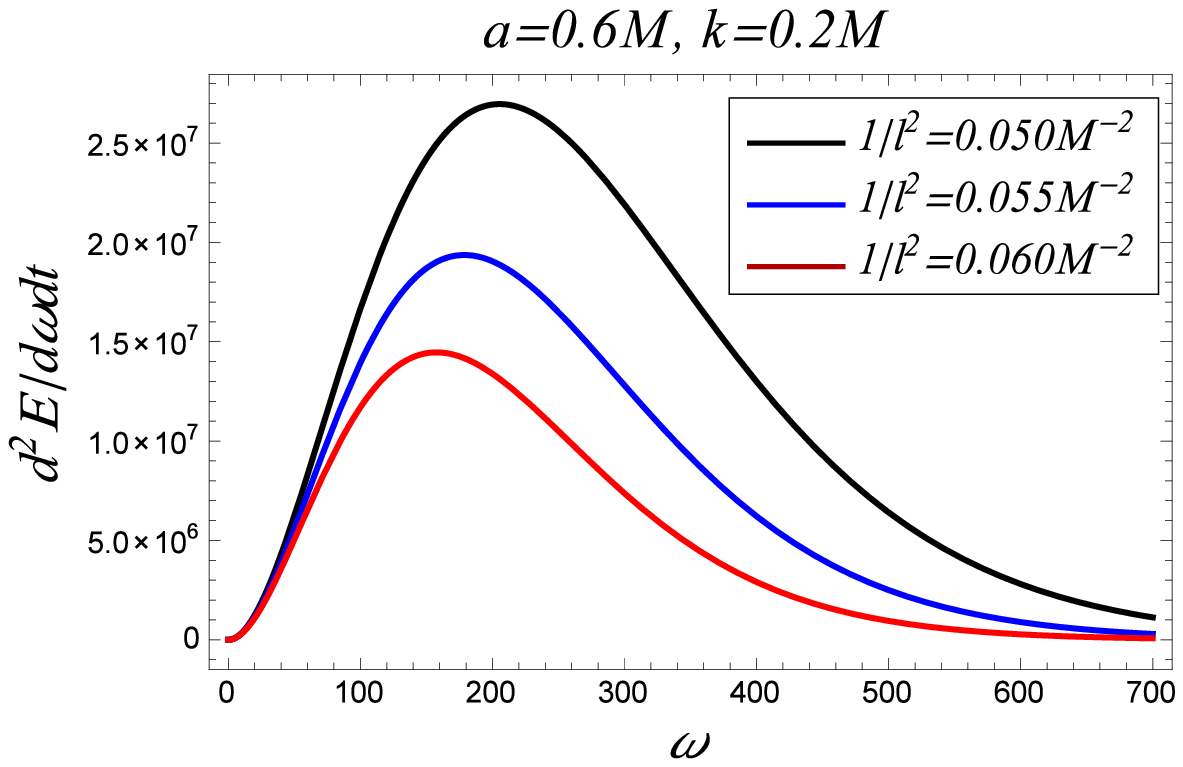} 
  	\end{tabular}
    \caption{\label{Nfig8} Plot showing the energy emission rate of the photon region .}
\end{figure*}
As a good approximation, we can use the the limiting constant value of the absorption cross section for a nearly spherically symmetric black hole as
\begin{equation}
\sigma_{\rm lim}\approx \pi R_s^2.  
\end{equation}
We can use such limiting value of the geometric cross-section in order to calculate the energy emission rate
\begin{eqnarray}
 \frac{d^2{E(\omega)}}{d\omega dt}=\frac{2\pi^2 R_s^2}{e^{\omega/T}-1}\omega^3,
\end{eqnarray}
where $\omega$ is the frequency of the emitted photon and the temperature $T$ of the nonsingular-AdS black hole in terms of the event horizon radius is expressed as
\begin{eqnarray}
 T=T_{\rm Kerr-AdS}-\frac{k\left(1-r_+^2/\l^2\right)}{4\pi r_+^2(r_+^2+a^2)},
\end{eqnarray}
where $T_{\rm Kerr-AdS}$ is the corresponding temperature for the Kerr-AdS black holes 
\begin{eqnarray}
 T_{\rm Kerr-AdS}=\frac{r_+^2 \left(1+3 r_+^2/l^2\right)-a^2 \left(1-r_+^2/l^2\right)}{4 \pi  r_+ \left(r_+^2+a^2\right)}.
\end{eqnarray}
  In Fig.~\ref{Nfig8}  we have plot the energy emission rate of the black hole  with $\omega$ for different values of curvature radius term. It is  shown in the figure that there exists a peak of energy emission rate, which decreases as we increase   $1/l^2$ term and  other black hole parameters.  

 \section{Conclusion}\label{sect6}
  In rotating Kerr spacetimes the shapes of black hole shadow get distorted mainly by the spin parameter $a$. In other theories of gravity $viz.$ the MGs, some other parameters may enhance or decrease the shadow of the black hole. In this paper, we have investigated the shadow of rotating regular black hole in the presence of a cosmological constant. The size of the black hole shadow and distortion parameter increases with the increasing values of  curvature radius parameter $1/l^2$. The black hole shadow appears smaller and more distorted in comparison to the regular and Kerr spacetime (cf. Fig.~\ref{Nfig3}). For a better understanding of black hole shadow, we have also calculated observables the shadow radius $R_s$ and the distortion parameter $\delta_s$. The observables $R_s$ and $\delta_s$ increases with the increasing values  and $1/l^2$ and the effective size of the shadow radius decreases with the free parameter $k$. From the study of energy emission rate we conclude that the electromagnetic spectrum emitted by the shadow region of regular black holes in AdS spacetime decreases for the increasing values of the parameter $1/l^2$.
 
 \section*{Acknowledgement}
The research of M. S. A. is supported by the National Postdoctoral Fellowship of the Science and Engineering Research Board (SERB), Department of Science and Technology (DST), Government of India, File No., PDF/2021/003491. B. P. S. would  like to thank Tulas Institute for providing research facilities.

\end{document}